\documentstyle[aps,preprint,amssymb,epsf]{revtex}
\begin{document}

\title{
 The Equation of State of Deconfined Matter at Finite Chemical Potential
 in a Quasiparticle Description}

\author{{\sc
 A.~Peshier$^{1,2}$, B.~K\"ampfer$^1$, G.~Soff$^2$}}

\address{
 $^1$Forschungszentrum Rossendorf, PF 510119, 01314 Dresden, Germany\\
 $^2$Institut f\"ur Theoretische Physik,
     Technische Universit\"at Dresden, 01062 Dresden, Germany}

\date{\today}

\maketitle

\begin{abstract}
 A quasiparticle description of the thermodynamics of deconfined matter,
 reproducing both the perturbative limit and nonperturbative lattice QCD
 data at finite temperature, is generalized to finite chemical potential.
 By a flow equation resulting from Maxwell's relation, the equation of state
 is extended from zero to non-zero quark densities.
 The impact of the massive strange flavor is considered and implications for
 cold, charge-neutral deconfined matter in $\beta$-equilibrium in compact
 stars are given.
 \\
 {\it Key Words:} deconfinement, equation of state, quark stars
 \\
 {\it PACS number(s):} 12.38.Mh, 31.15.Lc
\end{abstract}
\vspace*{5mm}

\section{Introduction}

 The equation of state (EoS) represents an important interrelation of
 state variables describing matter in local thermodynamical equilibrium.
 All microscopic characteristics are integrated out and only the macroscopic
 response to changes of the state variables as, e.\,g., the temperature $T$
 and the chemical potential $\mu$ are retained.
 Via the Gibbs equation, $e = T s + \mu n - p$, the energy density $e$ is
 related to the entropy and particle densities, $s = \partial p / \partial T$
 and $n = \partial p / \partial \mu$, respectively, and the pressure $p$.
 The thermodynamical potential $p(T, \mu)$ thus provides all information
 needed to evaluate, e.\,g., sequences of stellar equilibrium configurations
 by means of the Tolman-Oppenheimer-Volkov equation, or the evolution of the
 universe via Friedman's equations, or the dynamics of heavy-ion collisions
 within the framework of relativistic Euler equations.
 In the examples mentioned not only excited hadron matter is of relevance,
 rather, at sufficiently high density or temperature, a plasma state of
 deconfined quarks and gluons is the central issue.

 Within the quantum field theoretical standard model, the interaction of
 quarks and gluons is described by Quantum Chromodynamics (QCD).
 The challenge, therefore, consists in the derivation of the EoS of
 deconfined matter directly from QCD.
 In first attempts \cite{McLerran} the EoS of cold quark matter was
 derived perturbatively up to order ${\cal O}(g^4)$ in the coupling $g$.
 Finite temperatures have been considered, where the perturbative expansion
 is nowadays extended up to the order ${\cal O}(g^5)$ \cite{Zhai}.
 However, in the physically relevant region, due to the large coupling,
 perturbative methods seem basically to fail and, consequently,
 nonperturbative evaluations are needed.
 Present lattice QCD calculations can accomplish this task, and indeed the
 EoS of the pure gluon plasma \cite{EoS_g,Okamoto} and of systems containing
 two or four light dynamical quark flavors \cite{EoS_2f,EoS_4f,Karsch} are
 known reliably at finite temperature, and simulations with physical quark
 masses will possibly be available in the future.

 As a matter of fact, the current QCD lattice calculations are yet
 restricted to zero chemical potential $\mu$.
 Attempts to extend the lattice calculations systematically to non-zero
 chemical potential are under way \cite{non_zero}, but focus currently
 more on conceptional questions than on precise predictions as in the
 case of $\mu=0$.
 However, a detailed understanding of the EoS at non-vanishing net quark
 density is required for a variety of physical questions as, e.\,g., the
 structure of quark cores in massive neutron stars, the baryon contrast
 prior to cosmic confinement, or the evolution of the baryon charge in
 the mid-rapidity region of central heavy-ion collisions.

 Being aware of the urgent need for this EoS, we here suggest an
 approach, based on a quasiparticle description of quarks and gluons,
 to map available lattice data from $\mu=0$ to finite values of $\mu$
 and to small temperatures.
 Confidence in the model is gained by its quantitative agreement with the
 latest lattice data, which are analyzed in Sec.\ II, and by the smooth
 interpolation to the asymptotic limit. Discussed as well is the connection
 to the thermodynamic hard-thermal-loop resummation proposed recently
 \cite{ABS,BIR,HTLLW}, which provides a more rigorous motivation of our
 phenomenological approach.
 In Sec.\ III, we formulate the extension of the model to finite chemical
 potential, which does not require any further assumption. Although our
 approach is formulated in terms of gluon and quark quasiparticles and,
 hence, designed for `conventional' deconfined matter, there are formal
 indications of the existence of other QCD phases.
 In Sec.\ IV the quark-gluon plasma is considered with the massive
 strangeness degree of freedom appropriately taken into account. Despite
 of two free model parameters (which can be fixed when lattice data for
 the physical (2+1) flavor system will be available), our approach makes
 some definite statements about the EoS of the quark-gluon plasma. This
 is shown in a detailed study of the thermodynamics of $\beta$-stable
 deconfined matter, which allows us to draw conclusions about the bulk
 properties of pure quark stars.
 Finally, a summary and an outlook are given in Sec.\ V.

\section{The Quasiparticle model}
 The concept of describing an interacting system in terms of a system
 of quasiparticles which, as appropriate ground states, reflect relevant
 aspects of the interaction by effective properties, is successfully used
 in many fields of physics.
 For deconfined QCD plasmas, the development of the quasiparticle idea,
 starting with \cite{BLM,GS,PL}, ran parallel to the improvement of the
 lattice QCD calculations. These nonperturbative numerical simulations
 provide the testing ground for the quasiparticle concept which is a
 phenomenological approach.
 At the same time, by identifying the quasiparticles with the asymptotically
 relevant excitations \cite{PR}, these models yield a physically intuitive
 picture of the strongly interacting system.
 Referring to \cite{LevaiHeinz} for a recent survey of quasiparticle
 descriptions of finite-temperature lattice data, we focus in this work
 on the generalization of the approach to non-zero chemical potential.

\subsection{Formulation of the model}
 We consider an $SU(N_c)$ plasma of gluons, with $N_c=3$ for QCD, and
 $N_f$ quark flavors in thermodynamic equilibrium. Within the approach
 outlined below, the interacting plasma is described as a system of
 massive quasiparticles, a picture arising asymptotically from the
 in-medium properties of the constituents of the plasma.
 For thermal momenta $k \sim T$, the relevant modes, transversal gluons
 and quark-particle excitations, propagate predominantly on-shell with
 dispersion relations $\omega_i^2(k) \approx m_i^2 + k^2$ and
 \begin{equation}
  m_i^2
  =
  m_{0i}^2 + \Pi_i^* \, ,
  \label{m_eff}
 \end{equation}
 while the plasmon (longitudinal gluon) and plasmino (quark-hole) states
 are essentially unpopulated \cite{LeBellac}. Neglecting sub-leading
 effects, the medium contributions $\Pi_i^*$ are the leading order on-shell
 selfenergies.
 Depending on the coupling $G^2$, the temperature and chemical potential
 as well as the rest mass $m_{0i}$ (the latter vanishing for gluons), the
 $\Pi_i^*$ are given by the asymptotic values of the gauge independent
 hard-thermal/density-loop selfenergies \cite{LeBellac},
 \begin{eqnarray}
  \Pi_q^*
  &=&
  2\, \omega_{q0}\, (m_0 + \omega_{q0}) \, ,
  \quad
  \omega_{q0}^2
  =
  \frac{N_c^2-1}{16N_c}\, \left[ T^2+\frac{\mu_q^2}{\pi^2} \right] G^2 \, ,
  \nonumber \\
  \Pi_g^*
  &=&
  \frac16
  \left[
     \left( N_c+\frac12\, N_f \right) T^2
   + \frac{N_c}{2\pi^2} \sum_q \mu_q^2
  \right] G^2 \, .
 \label{Pi}
 \end{eqnarray}
 We emphasize that the contributions $\Pi_i^*$ are generated dynamically by
 the interaction within the medium, so the quantities $m_i$ in (\ref{m_eff})
 can be considered only as effective masses, with no ambiguities arising for
 the gauge boson quasiparticles.

 Generalizing the approach of Ref.~\cite{Gorenstein} to a finite chemical
 potential $\mu$ controlling a conserved particle number, the pressure of
 the system can be decomposed into the contributions of the quasiparticles
 and their mean field interaction $B$,
 \begin{equation}
  p(T,\mu; m_{0j}^2)
  =
  \sum_i p_i(T, \mu_i(\mu); m_i^2) - B(\Pi_j^*) \, ,
  \label{p_eff}
 \end{equation}
 where
 $p_i = \pm d_i\,T \int d^3k/(2\pi)^3\,\ln(1\pm\exp\{-(\omega_i-\mu_i)/T\})$
 is the pressure of an ideal gas of bosons or fermions with degeneracy
 $d_i$ and the state dependent effective masses (\ref{m_eff},\ref{Pi}).
 By the general stationarity of the thermodynamic potential $\Omega = -pV$
 under functional variation of the selfenergies \cite{LeeYang}, which in the
 present approach simplifies to $\partial p / \partial \Pi_j^* = 0$, $B$
 is related to the quasiparticle masses,
 \begin{equation}
   \frac{\partial B}{\partial \Pi_j^*}
   =
   \frac{\partial p_j(T,\mu_j; m_j^2)}{\partial m_j^2} \, .
  \label{stat}
 \end{equation}
 We remark that the stationarity of the quasiparticle pressure leads, by
 the Feynman-Hellmann relation, to a chiral condensate $\langle \bar\psi
 \psi \rangle \sim m_{0q}$, so the restoration of chiral symmetry of
 massless quark flavors in the deconfined phase is inherent in the approach.
 Furthermore, the stationarity implies that the entropy and the particle
 densities are given by the sum of the quasiparticle contributions,
 \begin{equation}
   s_i
   =
   \left.
    \frac{\partial p_i (T, \mu_i; m_i^2)}{\partial T}
   \right|_{m_i^2}
   \! , \quad
   n_i
   =
   \left.
    \frac{\partial p_i(T, \mu_i; m_i^2)}{\partial \mu_i}
   \right|_{m_i^2}
   \, .
  \label{s,n}
 \end{equation}
 This quasiparticle approach reproduces the leading-order perturbative
 results for arbitrary values of $N_c$ and $N_f$.
 Moreover, however, it represents a thermodynamically consistent effective
 resummation of the leading-order thermal contributions \cite{TFT}.
 As a nonperturbative approach, hence, it can be expected to be more
 appropriate in the large-coupling regime than perturbative predictions
 which, as truncated power expansions, fail to be reliable there.
 More specifically, the effective quasiparticle description will be an
 appropriate framework as long as the spectral properties of the relevant
 excitations do not differ qualitatively from their asymptotic form.
 Since propagating constituents in a plasma are, on general grounds, expected
 to possess a spectral representation with the strength predominately
 accumulated in a quasiparticle peak, our approach may be anticipated
 to be reasonable even close to the confinement transition.
 In a way, this expectation has some support from the hot $\phi^4$ theory
 where the selfconsistent resummation of superdaisy graphs leads to a
 similar `strict-quasiparticle' picture \cite{EL} as introduced above
 for QCD. By resumming the propagator beyond the leading-loop order, then,
 the quasiparticle properties were shown to be pronounced even for larger
 values of the coupling \cite{WangHeinz}.
 In a more direct way, the quasiparticle approach can be tested by comparison
 to the lattice data available for several QCD-model systems at finite
 temperature.

\subsection{Test of the model}
 To obtain the EoS as a function of the temperature, $G^2(T)$ has to be
 specified in Eq.~(\ref{Pi}). The quantity $G^2$ is to be considered as an
 {\em effective} coupling since it parameterizes all deviations of the
 exact spectral function from our `strict-quasiparticle' {\em ansatz}.
 On the other hand, to make contact to perturbation theory, it has to
 approach the running coupling in the asymptotic limit of large
 temperatures. In the form
 \begin{equation}
   G^2(T,\mu=0)
   =
   \frac{48\pi^2}
        {\left( 11 N_c - 2\, N_f \right)
         \ln\left(\displaystyle\frac{T+T_s}{T_c/\lambda}\right)^{\!\!2}} \, ,
   \label{G2}
 \end{equation}
 the shape of the exact spectral function is encoded in the single
 number $T_s/T_c$, whereas $T_c/\lambda$ is related to the QCD scale
 $\Lambda_{\rm QCD}$. While the physically intuitive ansatz (\ref{G2})
 turns out to be applicable to all systems considered in the following,
 other parameterizations are conceivable as well \cite{LevaiHeinz}.

\begin{figure}[hbt]
  \epsfysize 8cm
  \centerline{\epsffile{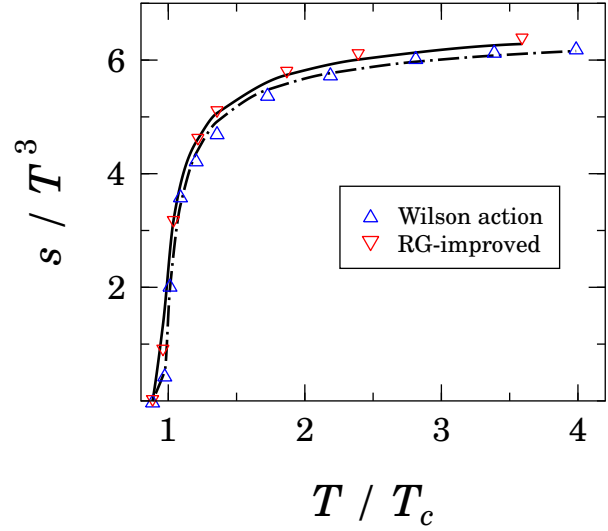}}
  \caption{The comparison of the quasiparticle entropy density with
           the lattice results, calculated with a standard Wilson action
           \protect\cite{EoS_g,Okamoto} and a renormalization-improved action
           \protect\cite{Okamoto}, of the pure gluon plasma.
           \label{F:SU3}}
\end{figure}

 As a first example, the quasiparticle model is applied to the case
 of the pure SU(3) gauge plasma.
 Recently, new lattice results calculated with a renormalization-improved
 action have been published \cite{Okamoto}, which are similar but by some
 3 $\cdots$ 4\% larger than the data \cite{EoS_g} obtained with a standard
 action. Consequently, the parameters fitting both data sets, as shown in
 Tab.\ 1, differ notably only in the number $d_g$ of the gluon quasiparticle
 degrees of freedom. The quantity $d_g$ is introduced here as fit parameter
 to account, on one hand, for possible finite size effects of the lattice
 data and, on the other hand, for residual sub-leading effects otherwise
 neglected in the quasiparticle ansatz. As expected, for both data sets the
 values of $d_g$ are close to $2(N_c^2-1) = 16$. As shown in Fig.~\ref{F:SU3}
 for the entropy, which is the quantity to adjust the parameters according
 to Eq.~(\ref{s,n}), the quasiparticle model reproduces the lattice results
 quantitatively. In fact, the deviations are smaller than the numerical
 uncertainty of the data.

\begin{figure}[hbt]
  \epsfysize 8cm
  \centerline{\epsffile{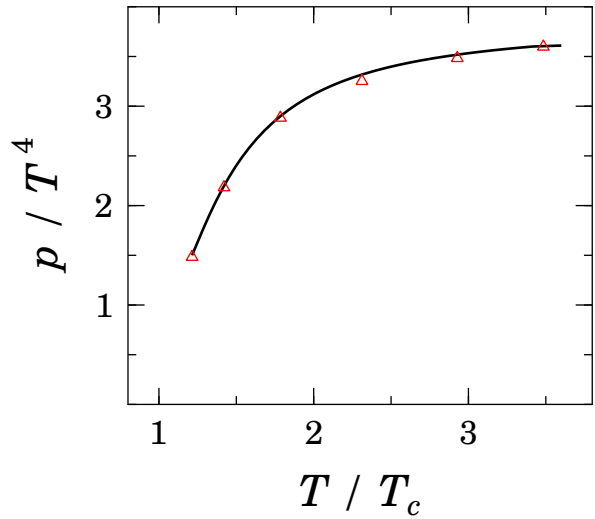}}
  \caption{The pressure in our model compared to the estimate of the
           continuum extrapolation of the lattice data \protect\cite{Karsch}
           for the light 2-flavor case.
           \label{F:Nf2}}
\end{figure}

 The next case analyzed is the system with two light quark flavors, with
 an estimate for the continuum extrapolation of the pressure given in
 \cite{Karsch}. To calculate the quasiparticle pressure (\ref{p_eff})
 for fixed values of the parameters, the interaction pressure $B$ is
 required as a function of the temperature. It can be obtained by the
 relation (\ref{stat}), with an integration constant $B_0 = B(T_0)$ adjusted
 to the lattice data, e.\,g., at the smallest temperature. The resulting
 agreement of the data and the model, with the parameters given in Tab.\ 1,
 is shown in Fig.~\ref{F:Nf2}. Note that the quark quasiparticle degeneracy
 is held fixed to $d_q=\frac{4N_c N_f}{2(N_c^2-1)}\, d_g$, as in the next
 example.

\begin{figure}[hbt]
  \epsfysize 8cm
  \centerline{\epsffile{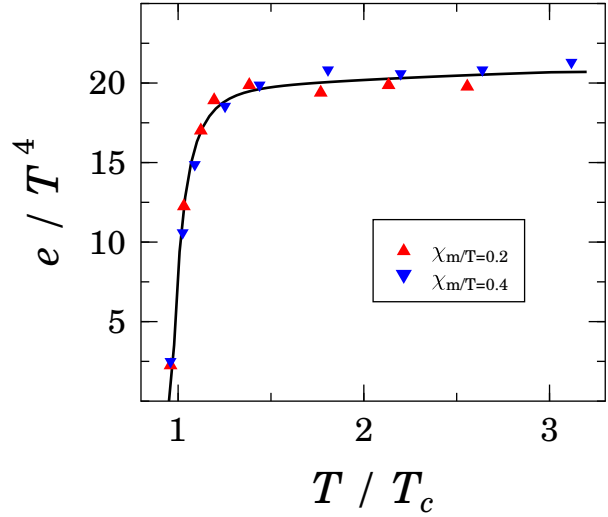}}
  \caption{The comparison of the energy density in our model with the
           chiral extrapolation of the lattice data \protect\cite{EoS_4f}
           of the 4-flavor case.
           \label{F:Nf4}}
\end{figure}

 The available lattice data \cite{EoS_4f} for the system with $N_f=4$ light
 flavors were obtained by the extrapolation to the chiral limit of lattice
 simulations of quarks with a ratio of mass to temperature of 0.2 and 0.4.
 Being compatible with each other, both data sets are reproduced by a single
 set of model parameters as shown in Fig.~\ref{F:Nf4}.
 So far, however, the data are not yet continuum extrapolated. Accordingly,
 the quasiparticle degeneracies are considerably larger than in the previous
 cases (cf.\ Tab.~1). Thus it is suggesting to interpret, to a large amount,
 the fit value of $d_g/16$ to be due to finite size effects which, in the
 free limit, are indeed found to have the same order of magnitude.
 Concerning the function $B(T)$ we mention that it turns out positive
 in the interval $1 \le T/T_c \le 2$, and negative and small at larger
 values of $T$ \cite{J_Phys}.

 We conclude this section with two remarks.
 As anticipated, the quasiparticle approach (being tested with available
 lattice data) is an appropriate description even close to the confinement
 transition. This provides some confidence in the applicability, in the
 nonperturbative regime, of the phenomenological quasiparticle resummation
 with the effects of finite spectral widths parameterized in an effective
 coupling.
 The quasiparticle approach is also supported by the more formal development
 of the thermodynamical resummation of hard-thermal loops, proposed in
 \cite{ABS} to improve the bare perturbation theory. As shown in \cite{BIR},
 the entropy of the plasma is dominated by the asymptotic behavior of those
 excitations which we start with as the relevant quasiparticles.
 Moreover, a direct hard-thermal-loop resummation of the pressure within
 the Luttinger-Ward formalism results in an expression similar to
 Eq.~(\ref{p_eff}), with $B$ corresponding in part to the so-called $\Phi$
 functional\footnote{
   Following these parallels further, we note that the hard-thermal-loop
   resummed entropy is independent of $\Phi$, as the quasiparticle entropy
   Eq.~(\protect\ref{s,n}) is independent of $B$. This feature, related to
   the underlying one-loop structure of the selfenergies in both cases,
   makes the entropy an interesting quantity, whereas the pressure contains
   the  full thermodynamical information.} \cite{HTLLW}.
 While this resummation, being more rigorous from a theoretical point of
 view and containing no fit parameters, reproduces the lattice data already
 qualitatively, we focus here on a phenomenological description, adjustable
 quantitatively to data.
 Nevertheless, the hard-thermal-loop resummation approaches yield further
 confidence to the quasiparticle model, also with regard to the extension
 to finite chemical potential as outlined subsequently.

\section{Extrapolating to finite chemical potential}
 Encouraged by the successful quasiparticle description of the $\mu=0$
 lattice data, the model is now applied to finite chemical potential.
 This extension requires no further assumptions and allows in a
 straightforward way to map the EoS at finite temperature and $\mu=0$
 into the $\mu\,T$ plane.
 This continuous mapping relying on quark and gluon quasiparticles,
 however, cannot provide information about other possible phases with
 a different (quasiparticle) structure, so the following consideration
 apply only to `conventional' deconfined matter.

 In general, the pressure is a potential of the state variables $T$ and
 $\mu$. As a direct consequence thereof, the Maxwell relation implies for
 the quasiparticle model
 \begin{equation}
   \sum_i
   \left[
     \frac{\partial n_i}{\partial m_i^2}\,
      \frac{\partial \Pi_i^*}{\partial T}
    -\frac{\partial s_i}{\partial m_i^2}\,
      \frac{\partial \Pi_i^*}{\partial \mu}
   \right]
   =
   0 \, ,
   \label{Max}
 \end{equation}
 which, formally, is the integrability condition for the function $B$
 defined by Eq.~(\ref{stat}). With $\Pi_i^*$ depending on $G^2$,
 Eq.~(\ref{Max}) represents a flow equation for the effective coupling,
 following directly from principles of thermodynamics.
 This flow equation is a quasilinear partial differential equation of
 the form
 \begin{equation}
   a_T\, \frac{\partial G^2}{\partial T}
   +
   a_\mu\, \frac{\partial G^2}{\partial \mu}
   =
   b \, ,
  \label{flow}
 \end{equation}
 with the coefficients $a_{T,\mu}$ and $b$ depending on $T$, $\mu$ and
 $G^2$. It is instructive to consider first the asymptotic limit, where
 $G^2 \rightarrow g^2 \ll 1$ and Eq.~(\ref{flow}) reduces to the
 homogeneous form
 \[
  \pi^2 \left( c T^2 + \frac{\mu^2}{\pi^2} \right)
  \frac1\mu\, \frac{\partial g^2}{\partial \mu}
  -
  \left( T^2 + \frac{\mu^2}{\pi^2} \right)
  \frac1T\, \frac{\partial g^2}{\partial T}
  =
  0 \, ,
 \]
 with $c = (4 N_c + 5 N_f)/(9 N_f)$.
 This equation has solutions $g^2 = const$ along the characteristics given
 by $c T^4 + 2 T^2 (\mu/\pi)^2 + (\mu/\pi)^4 = const.$ As a result of this
 elliptic flow, the renormalization scale $T_c/\lambda$ of $g^2(T,0)$
 determines the scale $T_c \pi c^{1/4}/\lambda$ of the $\mu$ dependent
 running coupling $g^2(0,\mu)$.

\begin{figure}[hbt]
  \epsfysize 8cm
  \centerline{\epsffile{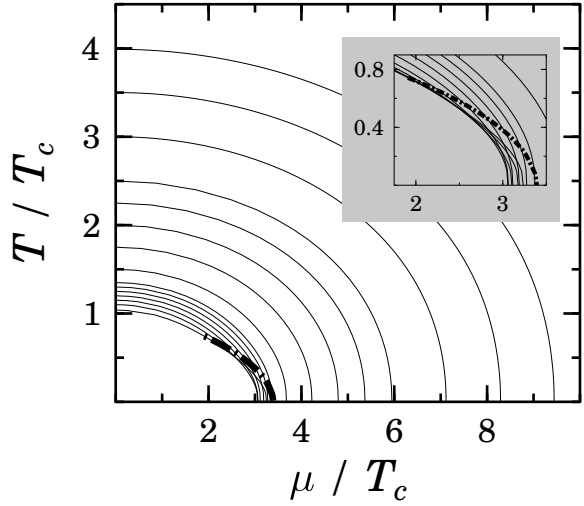}}
  \caption{The characteristics of the coupling flow equation
           (\protect\ref{flow}) for the QCD plasma with $N_f=4$ light
           flavors for which $G^2(T, \mu=0)$ is obtained from the lattice
           data \protect\cite{EoS_4f}.
           At leading order, the characteristics are curves of constant
           coupling strength.
           The region of intersecting characteristics is of no physical
           relevance since the pressure becomes negative in the region
           below the dash-dotted line (see text).
           \label{F:Nf4flow}}
\end{figure}

\begin{figure}[hbt]
  \epsfxsize 9cm
  \centerline{\epsffile{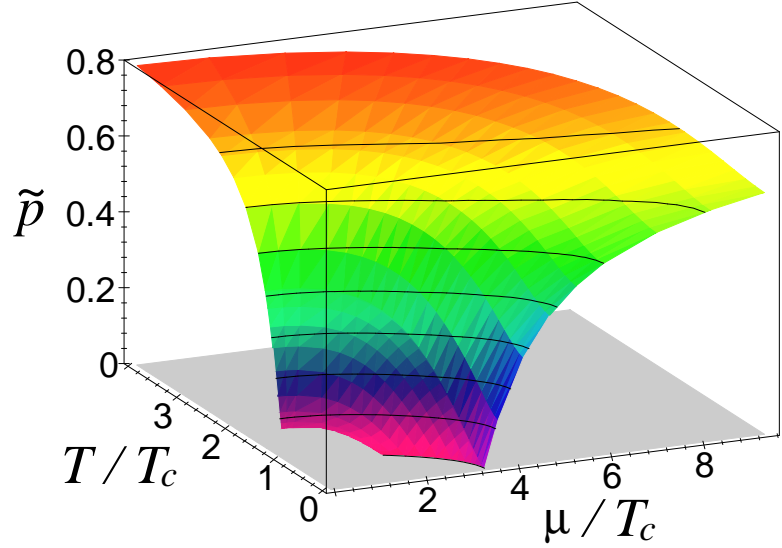}}
  \caption{The pressure of the $N_f=4$ plasma in the chiral limit as a
           function of $\mu$ and $T$, scaled by free limit.
           \label{F:Nf4p}}
\end{figure}

\begin{figure}[hbt]
  \epsfxsize 9cm
  \centerline{\epsffile{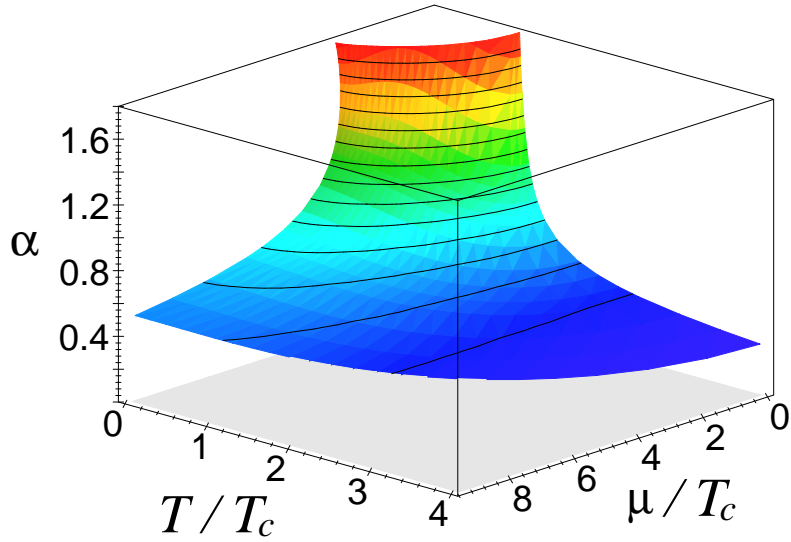}}
  \caption{The effective coupling $\alpha=G^2/(4\pi)$ of the $N_f=4$ plasma.
           \label{F:Nf4alph}}
\end{figure}

 The flow of the effective coupling is elliptic also in the nonperturbative
 regime, thus mapping the $\mu=0$ axis, where $G^2$ can be determined from
 lattice data, into the $\mu\,T$ plane.
 As a representative example, the nonperturbative flow of the coupling of
 the $N_f=4$ system is shown in Fig.~\ref{F:Nf4flow}. As to be expected,
 the characteristics emanating at $\mu=0$ in the deconfined phase do not
 attach a certain region, at small values of $T$ and $\mu$, of the phase
 diagram which is obviously related to the confinement phase. It is noted
 that at $\mu \mathrel{\rlap{\lower0.2em\hbox{$\sim$}}\raise0.2em\hbox{$<$}}
 3 T_c$ and small temperatures, indicated by intersecting characteristics,
 the solution of the flow equation is not unique. This ostensible ambiguity
 is, however, of no physical consequence, i.\,e., the approach is
 intrinsically consistent, since already outside that region the
 quasiparticle pressure (calculated along the characteristics) would become
 negative. This region of absolute thermodynamical instability, on the other
 hand, may be considered as an indication of a transition to a different
 phase, in fact already at some positive pressure.
 Whether this is the confined phase or, maybe already at such small chemical
 potential, a color-superconducting state (cf.\ \cite{color_super} and
 further references therein) remains, of course, a speculation within the
 present phenomenological approach.
 In any case, the pressure (see Fig.~\ref{F:Nf4p}) increases much slower
 with $\mu$ than with $T$, which makes the transition to another deconfined
 phase likely in that region.
 This behavior is mainly attributed to the fact that the effective coupling
 (see Fig.~\ref{F:Nf4alph}), as a function of increasing chemical potential,
 approaches the asymptotic limit more slowly than in the direction of
 increasing temperature.

 To conclude this example, we remark that the proposed extension of the
 lattice EoS to finite chemical potential, though model-dependent in using
 a quasiparticle picture to derive the flow equation (\ref{flow}), is based
 on {\em ab initio} calculable data. Since the model works perfectly at
 $\mu=0$ and, on the other hand, the convergence problem of perturbation
 theory and the concept to overcome it by resummation are not specific to
 $\mu=0$, we expect our extension of the data as reliable, at least
 semi-quantitatively.
 Similar flow equations can also be derived in more rigorous approaches,
 and on general grounds they are expected to be also of elliptic type.
 In this sense, the EoS at $\mu=0$ `knows' about $\mu \not= 0$.
 Nevertheless, as already mentioned, such continuous extensions do not
 access other possible QCD-phases with a different ground state; the actual
 EoS of strongly interacting matter is rather determined by thermodynamical
 stability between `competing' phases.

\section{Application of the model}

\subsection{Strangeness in matter}
 Although lattice simulations of the physically interesting case of two
 light and one medium-mass flavor are in progress \cite{Peikert}, reliable
 data of the thermodynamics of QCD with physical quark current masses,
 i.\,e.\ with $m_{0s} \sim 150$ MeV for the strange flavor, are still
 lacking to fix the quasiparticle parameters at $\mu=0$.
 Nevertheless, even when freely varying the parameters over large ranges,
 our model allows some predictions since the parameter dependence of the
 quasiparticle pressure $p^{qp}$ is weak due to the stationarity property
 Eq.~(\ref{stat}).
 The parameters are restricted to match $p^{qp}$ to the hadronic pressure
 $p^{had}$ at the confinement temperature which we assume to be $T_c=150\,$
 MeV. The uncertainty of $p^{had}(T_c,0) = 3.1\cdot10^8\,$ MeV$^4$, as
 estimated by a hadron resonance gas model, can be absorbed into the
 integration constant $B_0=B(T_c,0)$. This parameter, being related
 phenomenologically to the bag constant, is varied in the range 120\,MeV
 $\le B_0^{1/4} \le$ 180\,MeV. (Larger values of $B_0$ yield an increased
 ratio $e/T^4$ close to $T_c$, opposed to what is expected from available
 lattice data.)
 For the second independent parameter in $p^{qp}$, the range $3 \le \lambda
 \le 11$ is considered reasonably large, as suggested by the parameters of
 the analyzed lattice data in Tab.~1.
 The EoS for a representative choice of the parameters is provided in
 tabular form in \cite{WEB} for the plasma with vanishing net-strangeness,
 i.e.\ with the quark chemical potentials $\mu_s=0$ and $\mu_u=\mu_d=\mu$.
 With the effective coupling decreasing only logarithmically from $\alpha =
 G^2 / (4 \pi) \sim 1$ at small values of the temperature and the chemical
 potential, the medium contributions (\ref{Pi}) to the quasiparticle masses
 are large, of the order of $T$ or $\mu$. Hence, the quasiparticles are
 considerably under-saturated in phase space compared to the asymptotic
 limit, while the current mass of strange quarks turns out a less important
 suppression factor.

\subsection{Deconfined \boldmath$\beta$-stable strange matter}
\begin{figure}[hbt]
  \epsfysize 9cm
  \centerline{\epsffile{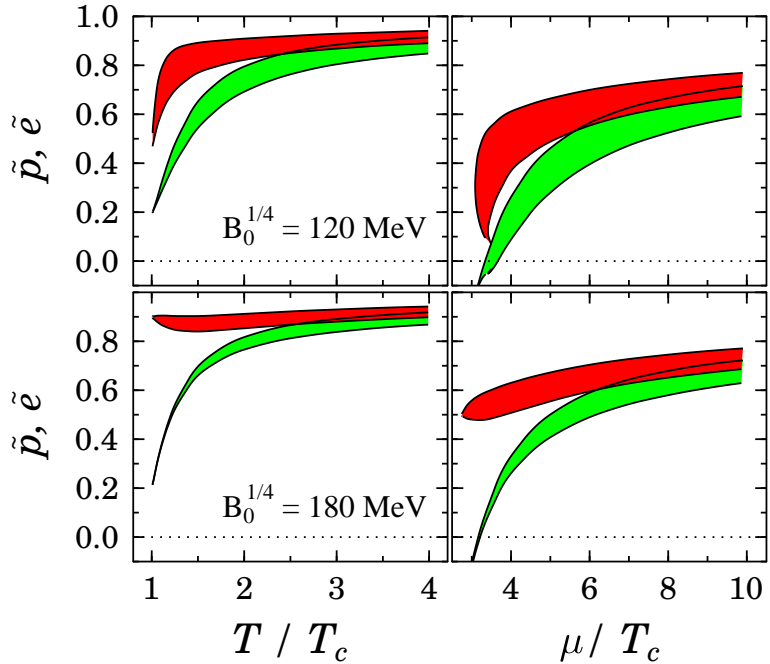}}
  \caption{Total pressure and energy density (lower and upper bands,
           respectively) of the charge-neutral quark-gluon plasma in
           $\beta$-equilibrium, scaled by the values of the free limit.
           The panels on the left (right) show the EoS at $\mu=0\;(T=0)$,
           for values of the model parameter $3 \le \lambda \le 11$
           (lower and upper line, respectively, of the shaded area),
           and $B_0^{1/4} = 120\,$MeV (top) and $B_0^{1/4} = 180\,$MeV
           (bottom).
           Non-unique values of the energy only occur in the unphysical
           region, where $p < 0$.
           \label{F:(2+1)eos}}
\end{figure}

 With the aim to study implications for quark matter stars, we consider in
 the following a plasma of gluons, quarks and leptons in $\beta$-equilibrium
 maintained by the reactions $d,s \leftrightarrow u + l + \bar\nu_l$, which
 imply the relations $\mu_d = \mu_s = \mu_u+\mu_l \equiv \mu$ among the
 chemical potentials.
 The lepton chemical potential $\mu_l$ is determined as a function of $T$
 and $\mu$ by the requirement of electrical charge neutrality.
 Without noticeable change of the result, the lepton pressure $p^l$ can be
 estimated by either the pressure of a free electron gas or, doubling the
 degrees of freedom, a gas of free electrons and muons.
 The pressure $p=p^{qp}+p^l$ and the resulting energy density at $\mu=0$
 are shown in the left panels of Fig.~\ref{F:(2+1)eos} for several values
 of the parameters $\lambda$ and $B_0$.
 Already at temperatures slightly above $T_c$, the scaled energy density
 reaches a saturation-like behavior at about 90\% of the asymptotic limit.
 This qualitative feature, known from the lattice simulations
 \cite{EoS_g,Okamoto,EoS_2f,EoS_4f,Karsch}, is to a large extent insensitive
 to the specific choice of the free parameters, while $B_0$ has a distinct
 impact on the latent heat.
 The resulting EoS at vanishing temperature is displayed in the right
 panels of Fig.~\ref{F:(2+1)eos}.
 In this regime, the asymptotic values are approached more slowly due to
 the less rapid decrease of the effective coupling with increasing values
 of $\mu$, similar to the $N_f=4$ plasma.
 For various sets of parameters, the EoS is available in tabular form
 \cite{WEB}.

\subsection{Pure quark stars}
 By the Tolman-Oppenheimer-Volkov (TOV) equation (cf.\ Eq.~(2.212)
 of Ref.~\cite{Glendenning}), sequences of hydrostatic equilibrium
 configurations of stars can be calculated, given the relation $e(p)$
 of the star matter. Since the bulk properties of the star are less
 dominated by the outermost shells, we consider the case of pure quark
 stars as a reasonable approximation for compact stars with a large quark
 matter core.
 For energy densities up to several times the nuclear saturation density
 and at temperatures less then some 10 MeV, the dependence $e(p)$ of
 $\beta$-stable quark matter, as estimated numerically by our quasiparticle
 model, can be parameterized by $e = 4 \tilde{B} + \tilde\alpha p$.
 While the naive bag model EoS had $\tilde\alpha = 3$, our EoS, with
 the considered choices of the model parameters, yields $3.1 \le \tilde
 \alpha \le 4.5$, indicating the nontrivial nature of the interaction.
 For parameters $\lambda \ge 5$, values of $\tilde{B}^{1/4} \ge 200\,$MeV
 are found, while only the extreme choice of $\lambda=3$ allows $\tilde{B}
 ^{1/4}$ as small as 180\,MeV.
 However, the value of $\tilde{B}$, i.\,e., the energy density at small
 pressure, has a strong impact on the star's mass and radius, obtained
 by integrating the TOV equation.

\begin{figure}[hbt]
  \epsfysize 8cm
  \centerline{\epsffile{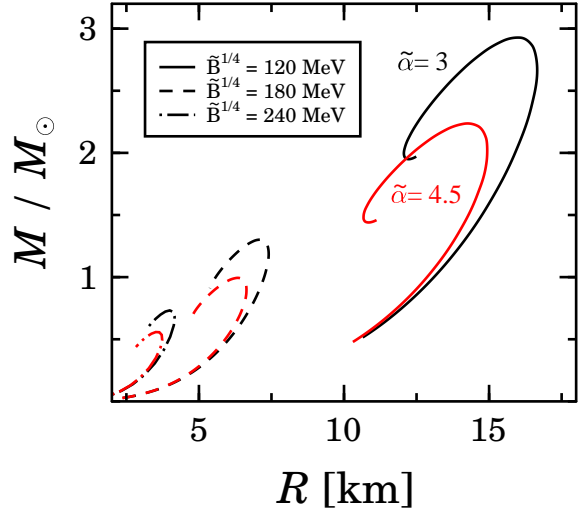}}
  \caption{The dependence of the mass of pure quark stars on the
           radius for several values of the parameters $\tilde \alpha$
           and $\tilde B^{1/4}$.
           \label{F:M(R)}}
\end{figure}

\begin{figure}[hbt]
  \epsfysize 8cm
  \centerline{\epsffile{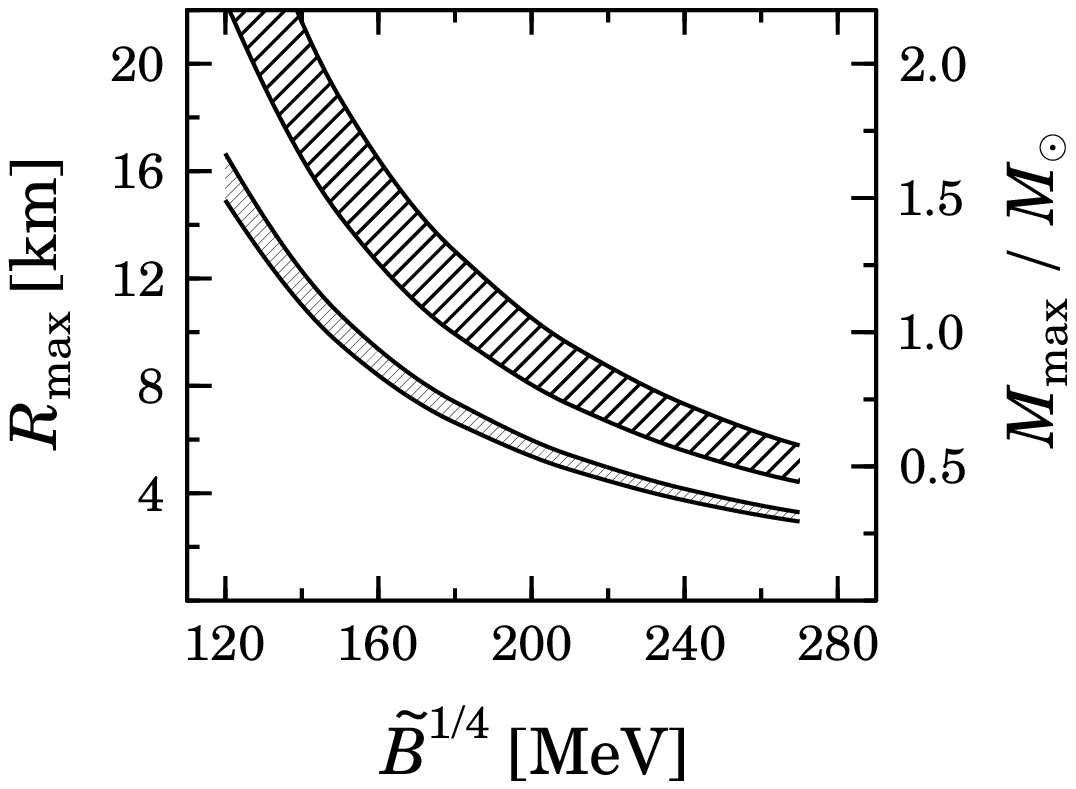}}
  \caption{The dependence of the maximum quark star mass
           (upper band and right scale)
           and the corresponding radius (lower band and left scale)
           on the parameter $\tilde B^{1/4}$.
           The upper (lower) boundary of the hatched bands are for
           $\tilde \alpha = 3$ (4.5).
           \label{F:MRmax}}
\end{figure}

 In Fig.~\ref{F:M(R)}, the mass as function the radius of pure quark
 stars is displayed for various values of the parameter $\tilde B$ and
 $\tilde \alpha$, while in Fig.~\ref{F:MRmax} the maximum mass and the
 corresponding radius are shown as function of $\tilde B^{1/4}$.
 For $\tilde B^{1/4}>200\,$MeV, as turned out to be the most likely range
 according to our analysis, the quark stars have masses less than one
 solar mass $M_\odot$. Compact quark stars with a larger mass can only
 exist for values of $\tilde{B}^{1/4} \le 180...200\,$MeV, depending only
 marginally on $\tilde \alpha$.
 We hence conclude that compact stars with masses $M \sim M_\odot$ and
 radii $R \sim 10\,$km are unlikely to be composed mainly of deconfined
 matter in $\beta$-equilibrium, irrespective of the uncertainty of the
 model parameters of our approach.

\section{Summary and outlook}
 In summary we have generalized a thermodynamic quasiparticle description
 of deconfined matter to finite chemical potential $\mu$ not yet accessible
 by present lattice calculations.
 Of central importance to the model is the effective coupling $G^2(T,\mu)$
 which can be obtained at $\mu=0$ from available lattice data, proving at
 the same time the applicability of the quasiparticle picture even close
 to the confinement transition.
 At $\mu \ne 0$, $G^2$ is determined by a flow equation following directly
 from the Maxwell relation of thermodynamics.
 As a result of the elliptic flow, the basic features of the EoS at $\mu=0$,
 namely the nonperturbative behavior near confinement and the asymptotic
 behavior, are mapped into the $\mu\,T$ plane as exemplified by the $N_f=4$
 flavor system studied on the lattice at $\mu = 0$.

 An important consequence of the quasiparticle approach is the relation
 of the `critical' values of temperature and chemical potential.
 For deconfinement matter with physical quark masses, this fact leads to
 the implication that pure quark stars are less massive and, hence, more
 distinct in the bulk properties to neutron stars than estimated by other
 approaches with an {\it ad hoc} choice of small values of the bag constant.

 Further applications of the quasiparticle model, in connection with
 realistic thermodynamical descriptions of other QCD phases, is desirable.
 In particular, in the cold and dense regime where a color-superconducting
 state is expected to be important, the quasiparticle EoS can provide one
 part of the information about the phase boundary.
 With regard to the on-going heavy-ion program at the relativistic heavy-ion
 collider RHIC at Brookhaven National Laboratory, we finally would like to
 point out the applicability of our quasiparticle EoS as an input to
 hydro-relativistic simulations. For a realistic description of the (near-to)
 equilibrium states of a heavy-ion collision, detailed knowledge about the
 EoS at non-zero baryon densities is required.
 Since the results of the quasiparticle model differ considerably from that
 of simpler approaches as the bag model (e.\,g., in the reduced latent heat
 and the saturation-like behavior of the energy density at about 90\% of
 the free limit), dynamical effects of a realistic EoS may even lead to
 observable signals of the experimental formation of the quark-gluon plasma.

 {\bf Acknowledgments:}
 We are grateful to E.~Grosse and F.~Thielemann for initiating the present
 work. The stimulating interest of O.\,P.~Pavlenko is acknowledged.
 The work is supported by BMBF 06DR829/1.

\begin{table}
  \begin{tabular}{|l||c|c|c|}
                                & $\lambda$ & $T_s/T_c$ & $d_g$ 
  \\ \hline \hline
  SU(3) \protect\cite{EoS_g}    &   5.02    &  -0.75    & 16.9
  \\ \hline
  SU(3) \protect\cite{Okamoto}  &   4.83    &  -0.72    & 17.5
  \\ \hline
  $N_f=2$ \protect\cite{Karsch} &   10.2    &  -1.00    & 17.0
  \\ \hline
  $N_f=4$ \protect\cite{EoS_4f} &   6.59    &  -0.80    & 20.6
  \end{tabular}
  \vspace*{3mm}
  \caption{Quasiparticle-model parameters adjusted to the lattice data.
           For the systems containing quarks, the quark degeneracy is
           fixed to $d_q=\frac{4N_c N_f} {2(N_c^2-1)}\, d_g$.}
\end{table}

\newpage

\begin{thebibliography}{99}
\bibitem{McLerran}
   B.\,A.~Freedman, L.\,D.~McLerran,
   Phys.\ Rev.\ D16, 1130, 1147, 1169 (1977)
\bibitem{Zhai}
   C.~Zhai, B.~Kastening,
   Phys.\ Rev.\ D52, 7232 (1995)
\bibitem{EoS_g}
   G.~Boyd, J.~Engels, F.~Karsch, E.~Laermann, C.~Legeland,
   M.~L\"utgemeier, B.~Petersson,
   Nucl.\ Phys.\ B469, 419 (1996)
\bibitem{Okamoto}
   M.~Okamoto et al.\ (CP-PACS collaboration), hep-lat/9905005
\bibitem{EoS_2f}
   C.~Bernard et al.,
   Phys.\ Rev.\ D55, 6861 (1997)
\bibitem{EoS_4f}
   J.~Engels, R.~Joswig, F.~Karsch, E.~Laermann, M.\ L\"utgemeier,
   B.~Petersson,
   Phys.\ Lett.\ B396, 210 (1997)
\bibitem{Karsch}
   F.~Karsch, hep-lat/9909006
\bibitem{non_zero}
   J.~Engels, O.~Kaczmarek, F.~Karsch, E.~Laermann, hep-lat/9903030
\bibitem{ABS}
   J.\,O.~Anderson, E.~Braaten, M.~Strickland, Phys.\ Rev.\ Lett.\
   83, 2139 (1999), hep-ph/9905337
\bibitem{BIR}
   J.\,P.~Blaizot, E.~Iancu, A.~Rebhan, Phys.\ Rev.\ Lett.\ 83, 2906 (1999),
   hep-ph/9910309
\bibitem{HTLLW}
   A.~Peshier, hep-ph/9910451
\bibitem{BLM}
   T.\,S.~Bir\'o, P.~L\'evai, B.~M\"uller,
   Phys.\ Rev.\ D42, 3078 (1990)
\bibitem{GS}
   V.~Goloviznin, H.~Satz,
   Z.\ Phys.\ C57, 671 (1993)
\bibitem{PL}
   A.~Peshier, B.~K\"ampfer, G.~Soff, O.\,P.~Pavlenko,
   Phys.~Lett.~B337, 235 (1994)
\bibitem{PR}
   A.~Peshier, B.~K\"ampfer, O.\,P.~Pavlenko, G.~Soff,
   Phys.\ Rev.\ D54, 2399 (1996)
\bibitem{LevaiHeinz}
   P.~Levai, U.~Heinz,
   Phys.\ Rev.\ C57, 1879 (1998)
\bibitem{LeBellac}
   M.~Le Bellac, {\em Thermal Field Theory},
   Cambridge University Press, Cambridge (1996)
\bibitem{Gorenstein}
   M.\,I.~Gorenstein, S.\,N.~Yang,
   Phys.\ Rev.\ D52, 5206 (1995)
\bibitem{LeeYang}
   T.\,D.~Lee, C.\,N.~Yang,
   Phys.\ Rev.\ 117, 22 (1960)
\bibitem{TFT}
   A.~Peshier,
   TFT'98 Proceedings, hep-ph/9809379
\bibitem{EL}
   A.~Peshier, B.~K\"ampfer, G.~Soff, O.\,P.~Pavlenko,
   Europhys.\ Lett.\ 43, 381 (1998)
\bibitem{WangHeinz}
   E.~Wang, U.~Heinz,
   Phys.\ Rev.\ D53, 899 (1996)
\bibitem{J_Phys}
   B.~K\"ampfer, O.\,P.~Pavlenko, A.~Peshier, M.~Hentschel, G.~Soff,
   J. Phys. G23, 2001c (1997)
\bibitem{color_super}
   R.\,D.~Pisarski, D.\,H.~Rischke,
   Phys.\ Rev.\ Lett.\ 83, 37 (1999)
\bibitem{Peikert}
   A.~Peikert, F.~Karsch, E.~Laermann, hep-lat/9909116
\bibitem{WEB}
   http://www.fz-rossendorf.de/FWK/MITARB/Peshier/EoS.html
\bibitem{Glendenning}
   N.\,K.~Glendenning,
   {\em Compact stars}, Springer Verlag, New York (1997)
\end{thebibliography}
\end{document}